\documentclass[12pt,a4paper]{article}

\usepackage{graphicx}
\usepackage {epsfig}
\usepackage {epstopdf}

\begin{document}
\textwidth=135mm
 \textheight=200mm
\begin{center}
{\bfseries Dynamical Schwinger effect: Properties of the $e^{+}e^{-}$ plasma 
created from vacuum in strong laser fields
}
\vskip 5mm
D. Blaschke$^{\dag,\ddag}$, L.~Juchnowski\footnote{E-mail: lukasz.juchnowski@ift.uni.wroc.pl}$^\ddag$, A.~Panferov$^\star$, 
S.~Smolyansky$^\star$

\vskip 5mm
{\small {\it $^\dag$ Bogoliubov Laboratory for Theoretical Physics, JINR Dubna, Russia}} \\
{\small {\it $^\ddag$ Institute for Theoretical Physics, University of Wroclaw, Poland}}
\\
{\small {\it $^\star$ Department of Physics, Saratov State University, Russia}}
\end{center}
\vskip 5mm
\centerline{\bf Abstract}
We study the dynamical Schwinger effect in the vacuum excitation of the electron-positron 
plasma under action of a "laser pulse" of the simplest configuration: a linearly polarized,
time-dependent and spatially homogeneous electric field. 
Methodical basis of this investigation is the kinetic equation which is an exact consequence 
of the basic equations of motion of QED in the considered field model. 
In the present work we investigate some features of the residual electron-positron plasma 
and the transient process of its formation.

\vskip 10mm
\section{\label{sec:intro}Introduction}
At the present time it is well known that the electron-positron plasma (EPP) excited from 
vacuum passes through three stages of its evolution: 
the quasiparticle stage in the period of the laser pulse action, 
the transient period of the EPP transmutation and 
the final residual EPP (REPP) in the out-state (particles in this state are on their mass 
shell). 
The quasiparticle period of the EPP evolution was investigated in detail in the work 
\cite{001}. 
In the present work we will study the transient period of the EPP evolution and its final 
state, the REPP. 
As a rule, the density of the quasiparticle EPP is more larger than that of REPP. 
However, all three stages give a contribution to the different observable effects as, e.g., 
the emission of annihilation photons from the focal spot of the colliding laser beams 
\cite{002a,002}. 
It is rather difficult to estimate the contributions of the different stages in formation 
of observable effects. 
Therefore the detailed investigation of each period of the EPP evolution is an important 
problem.

In this work we restrict ourselves to the consideration of the simplest model for the 
external ("laser") field: 
the linearly polarized, time-dependent and spatially homogeneous electric field pulse. 
It is assumed that such a field acts in the focal spot of counter-propagating laser beams 
in the optical or X-ray ranges.

Our investigations of the dynamical Schwinger effect in such fields are based on numerical 
solutions of the kinetic equation (KE) of non-Markovian type which is a nonperturbative 
consequence of the basic equations of motion of QED. 
This approach allows to investigate the behaviour of the EPP in both cases, the tunneling 
and the multiphoton mechanisms of EPP excitation.

\section{The kinetic equation in an external \\ time-dependent field}
The KE for the (quasi-)particle distribution function can be derived from the Dirac 
equation by a canonical time-dependent Bogoliubov transformation \cite{Blaschke}. 
This method is valid only in a spatially-uniform time-dependent field. 
In the case of a linearly polarized electric field with the vector potential 
$A^{\mu}(t) = (0,0,0,A(t))$  (Hamiltonian gauge) we obtain a non-Markovian integro-differential 
collisionless KE  
\begin{equation}\label{integro}
\dot f(\mathbf{p},t) = \frac{1}{2} \lambda(\mathbf{p},t)
\int\limits^t_{t_0} dt^{\prime} \lambda(\mathbf{p},t^{\prime})
[1-2f(\mathbf{p},t^{\prime})]\cos\theta(t,t^{\prime}),
\end{equation}
where
\begin{equation}\label{lambda}
\lambda(\mathbf{p},t) = e E(t)\varepsilon_{\bot}/\varepsilon^{2}(\mathbf{p},t),
\qquad   \theta(t,t^{'}) = 2 \int^t_{t^{`}} d\tau \, \varepsilon (\mathbf{p} ,\tau),
\end{equation}
with the transversal energy $\varepsilon_{\bot} = \sqrt{m^2 + p_{\bot}^2}$ 
and the quasienergy 
\begin{equation}\label{energy}
\varepsilon (\mathbf{p} ,t) = \sqrt{\varepsilon_{\bot}^2 + (p_{\parallel} + A(t))^2}.
\end{equation}

Here $\lambda$ is the amplitude of vacuum transitions and $\theta$ is a dynamical phase 
which describes the vacuum oscillations (Zitterbewegung) with a frequency of the energy 
gap $2\varepsilon (\mathbf{p} ,t)$.   
The equation contains two characteristic time scales. 
The slower scale is external field period and the faster one is given by the Compton time 
$\tau_C = 2\pi/m$.
The KE~(\ref{integro}) is equivalent to a system of ordinary differential equations (ODE's)
\begin{eqnarray}
\label{ode}
    \dot{f} = \frac12\, \lambda u, \qquad
\dot{u} = \lambda \, (1-2f) - 2 \varepsilon\, v, \qquad \dot{v}&=& 2 \varepsilon u .
\end{eqnarray} 

The total particle number density is defined by 
\begin{equation}\label{density}
n(t) = g \int \frac{d\vec{p}}{(2\pi)^3}f(\vec{p},t) , \nonumber
\end{equation}
where g = 4 is the degeneracy factor due to spin and charge degrees of freedom. 
The system of ODE's describes transitions between the lower and the upper continuum of the 
energy spectrum. 
These processes resemble interband transitions in solid state physics. 
Moreover, for excitation energies smaller than the energy gap a virtual state can be created.

We solve the KE~(\ref{integro}) numerically for the following field shapes 
\begin{equation} \label{field1}
E(t) = E_0 \cosh^{-2}(t/T), \qquad  A(t)= {TE_0} \tanh(t/T),
\end{equation}
which is the Eckart field, and \cite{004}
\begin{eqnarray} \label{field2} 
E(t) & = & E_0  \cos{ (\omega t+\phi) }\ e^{-t^2/2\tau^2 }, \\
A(t) &=& -\sqrt{\frac{\pi}{8}} E_0\tau \exp{(-\sigma^2/2+i\phi)} 
\mbox{erf} \left(\frac{t}{\sqrt{2}\tau} -i\frac{\sigma}{\sqrt{2}}\right) + c.c. , \nonumber
\end{eqnarray}
where $\sigma = \omega \tau$.
Equations (\ref{field1}), (\ref{field2}) model standing waves created by the two 
counter-propagating laser beams.

\section{Transient process and REPP forming}
\begin{figure}[!h]
  \begin{tabular}{cc}
  \includegraphics[angle=0,width=0.4 \linewidth]{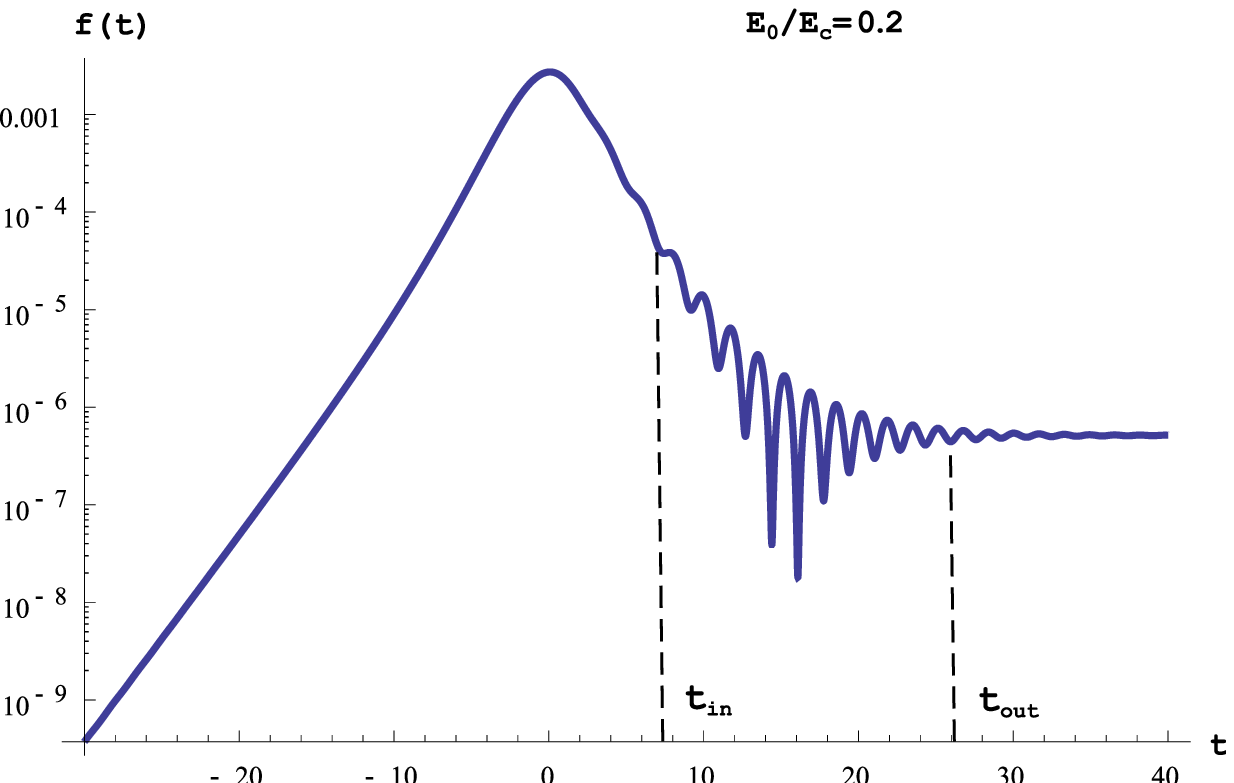}
  &
  \includegraphics[angle=0,width=0.4 \linewidth]{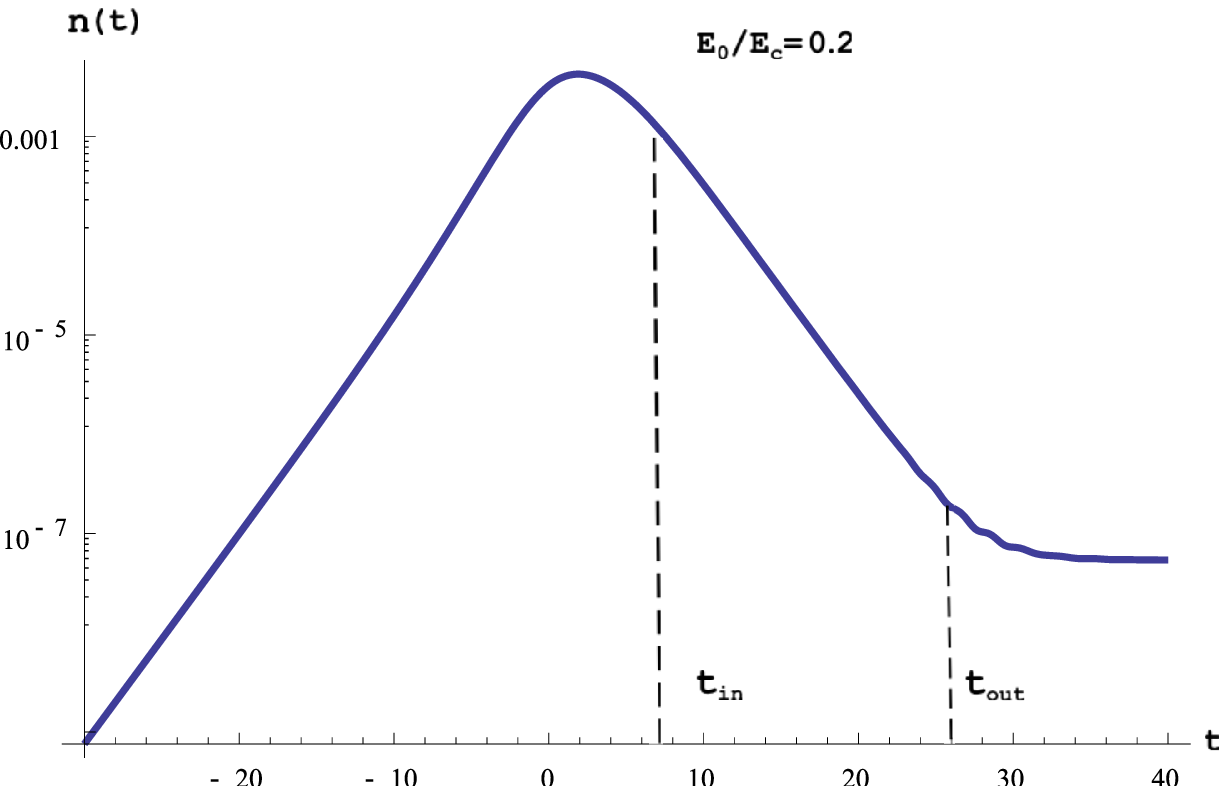}
  \end{tabular}
 \caption{Distribution function (left panel) and pair density (right panel) for the Eckart 
 field pulse $E(t) = E_0 \cosh^{-2}(t/T), \quad T = 0.02\mbox{nm}$.}
 \label{fig:eckertfield}
 \end{figure}

\begin{figure}[!h]
\begin{tabular}{cc}
\includegraphics[angle=0,width=0.4 \linewidth]{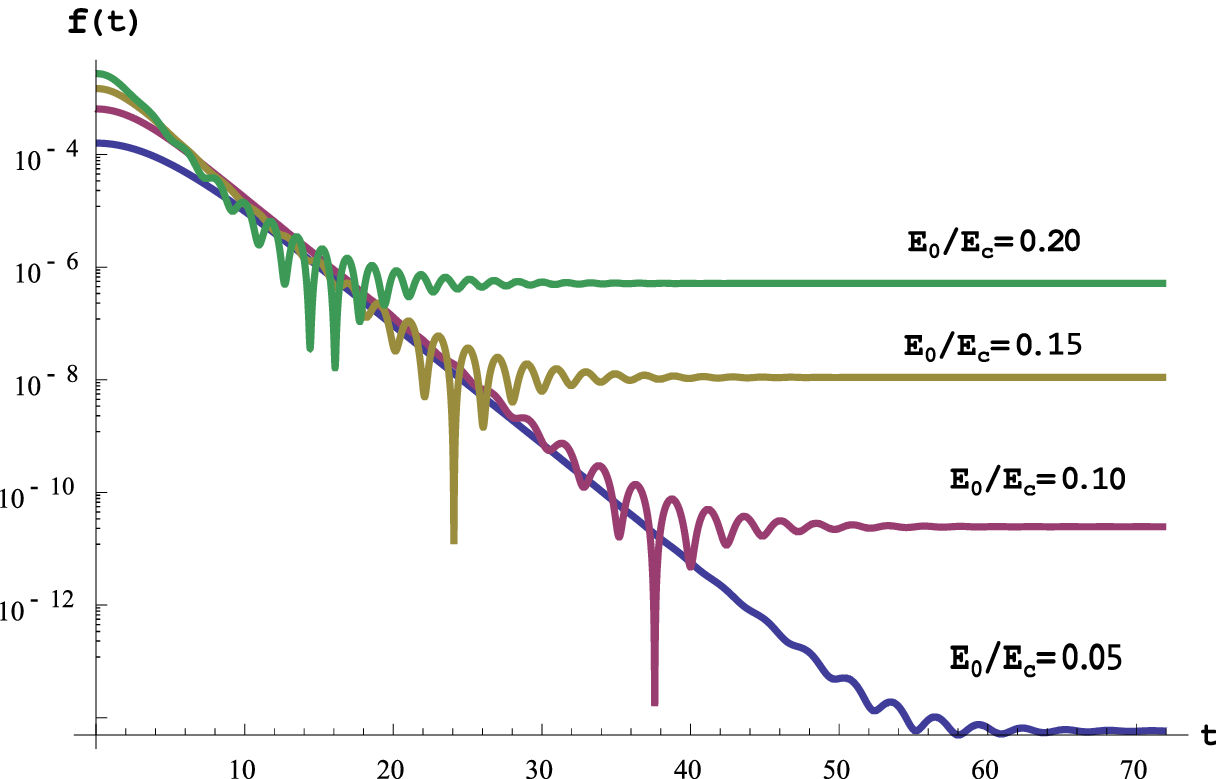}
&
\includegraphics[angle=0,width=0.4 \linewidth]{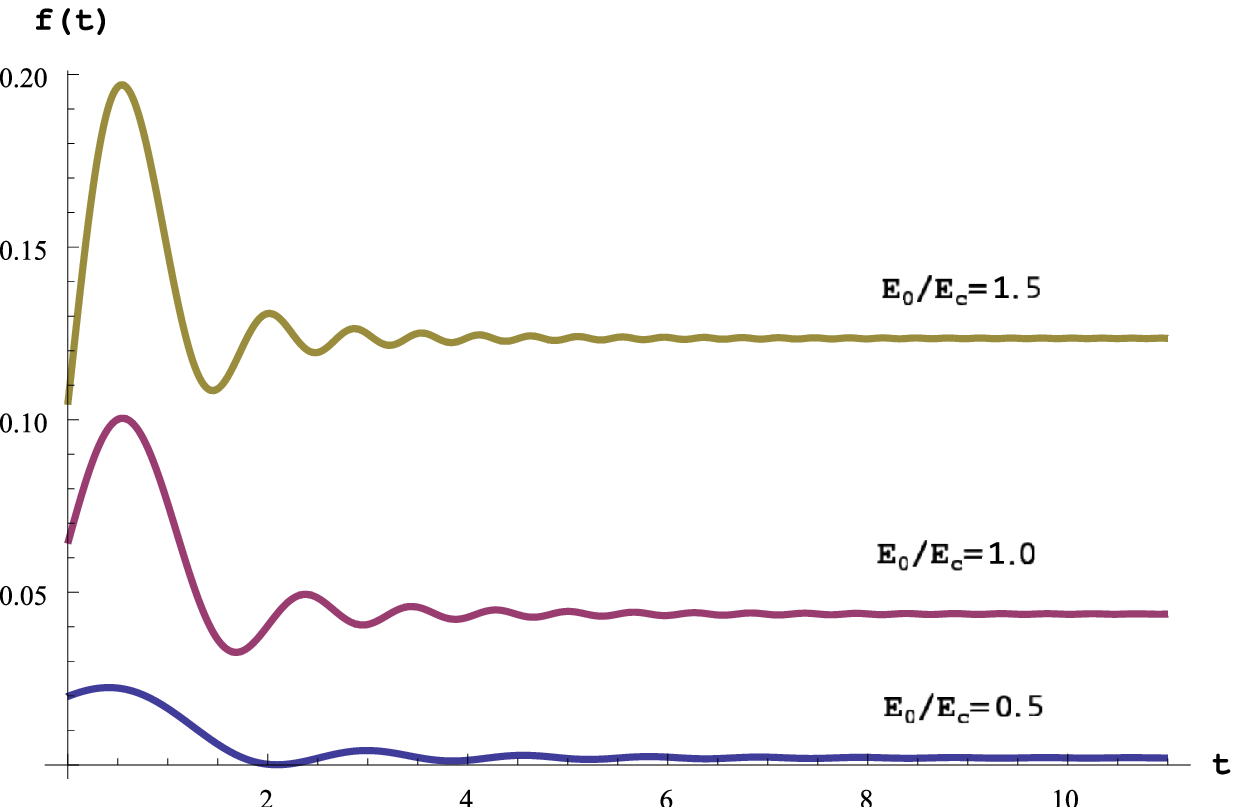}
\end{tabular}
\caption{Temporal behaviour of the maximum of the distribution function for different 
field strengths $E_0$. All other parameters as in Fig.~\ref{fig:eckertfield}.}
\label{fig:eckertfield2}
\end{figure}

\begin{figure}[h!]
\begin{tabular}{cc}
\includegraphics[angle=0,width=0.4 \linewidth]{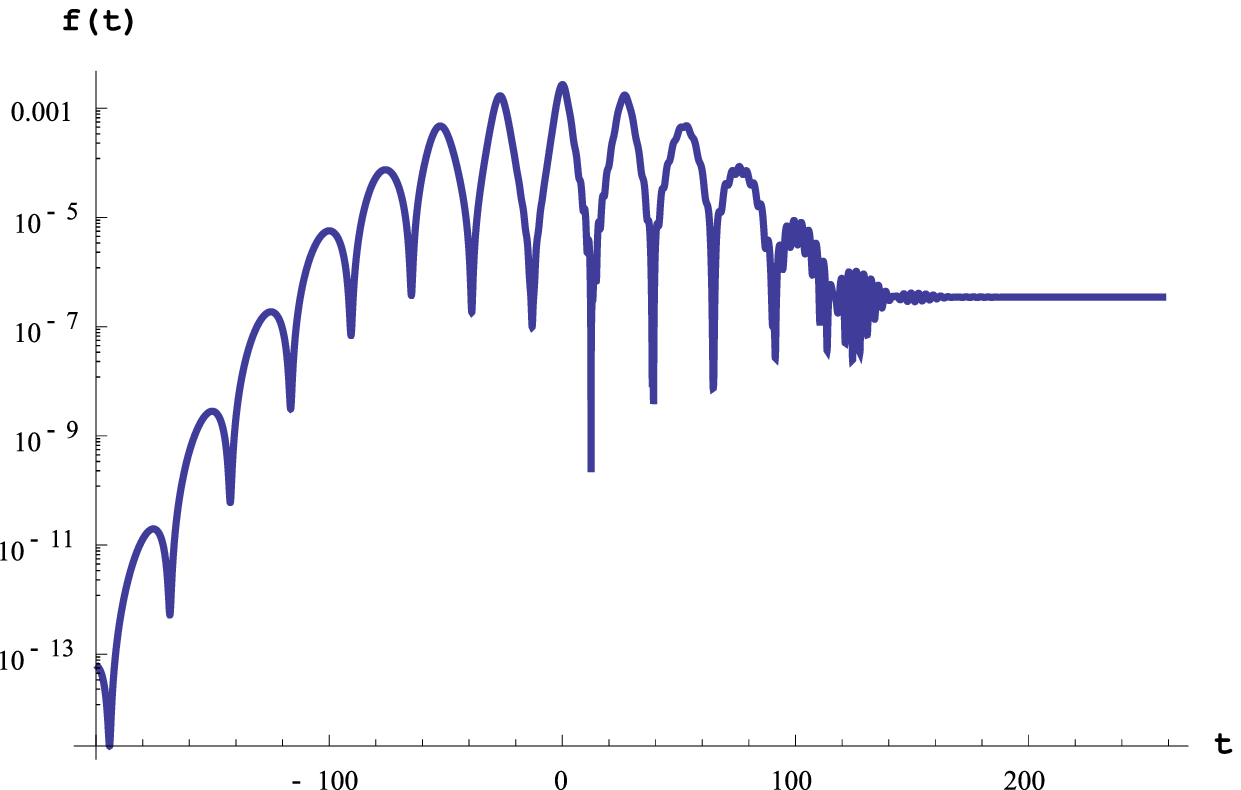}
&
\includegraphics[angle=0,width=0.4 \linewidth]{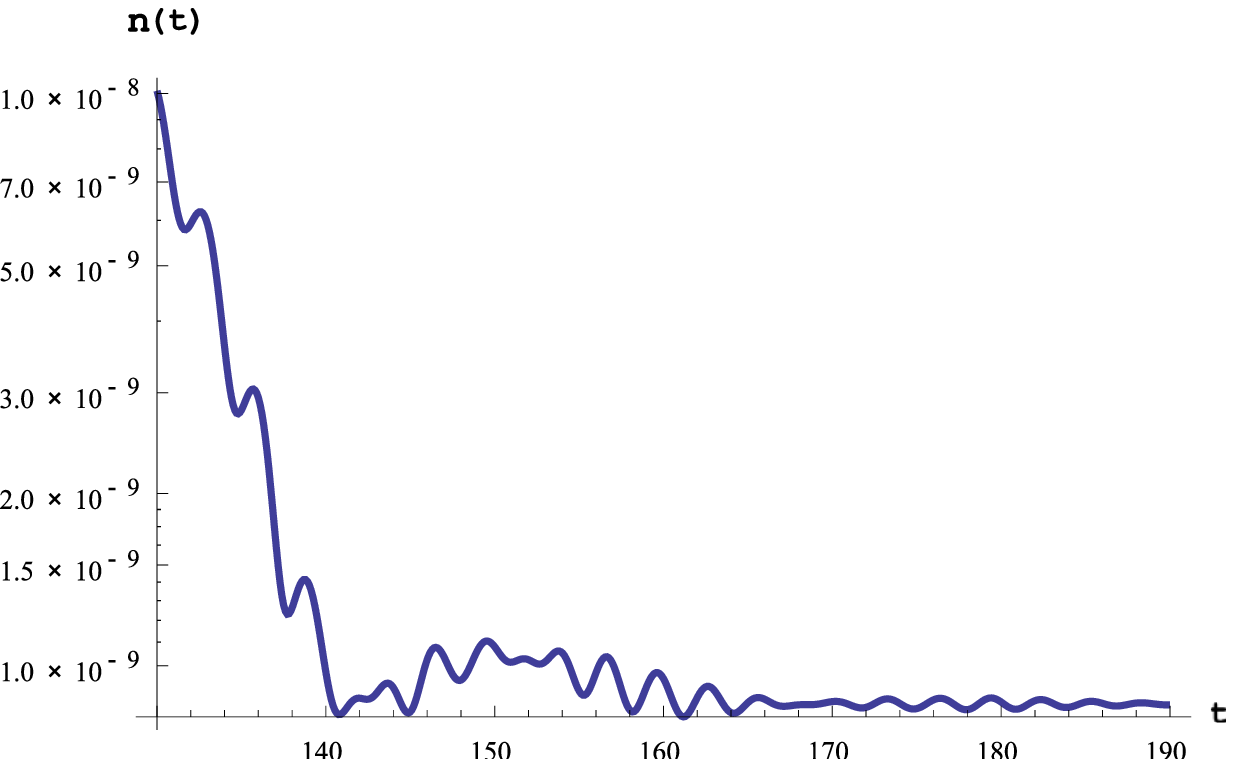}
\end{tabular}
\caption{Same as Fig.~\ref{fig:eckertfield} for the Gaussian modulated periodic field
$E(t)=E_0\cos{(\omega t+\phi)}\ e^{-t^2/2\tau^2 }, \quad \phi = 0,\quad 
\sigma = \omega\tau = 5$.}
\label{fig:gaussfield}
\end{figure}

\begin{figure}[!h]
\begin{tabular}{cc}
\includegraphics[angle=0,width=0.40 \linewidth]{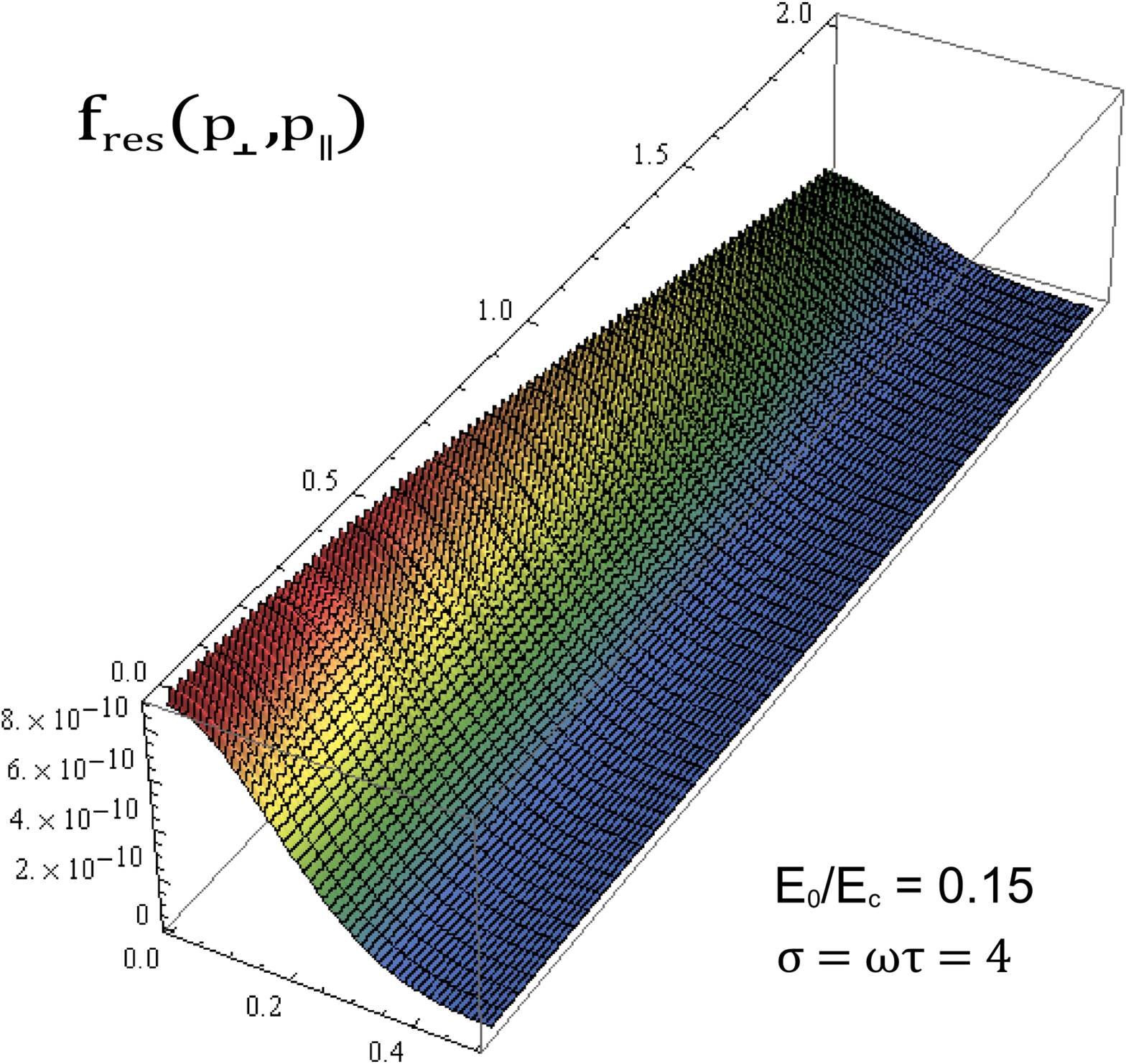}
&
\includegraphics[angle=0,width=0.40 \linewidth]{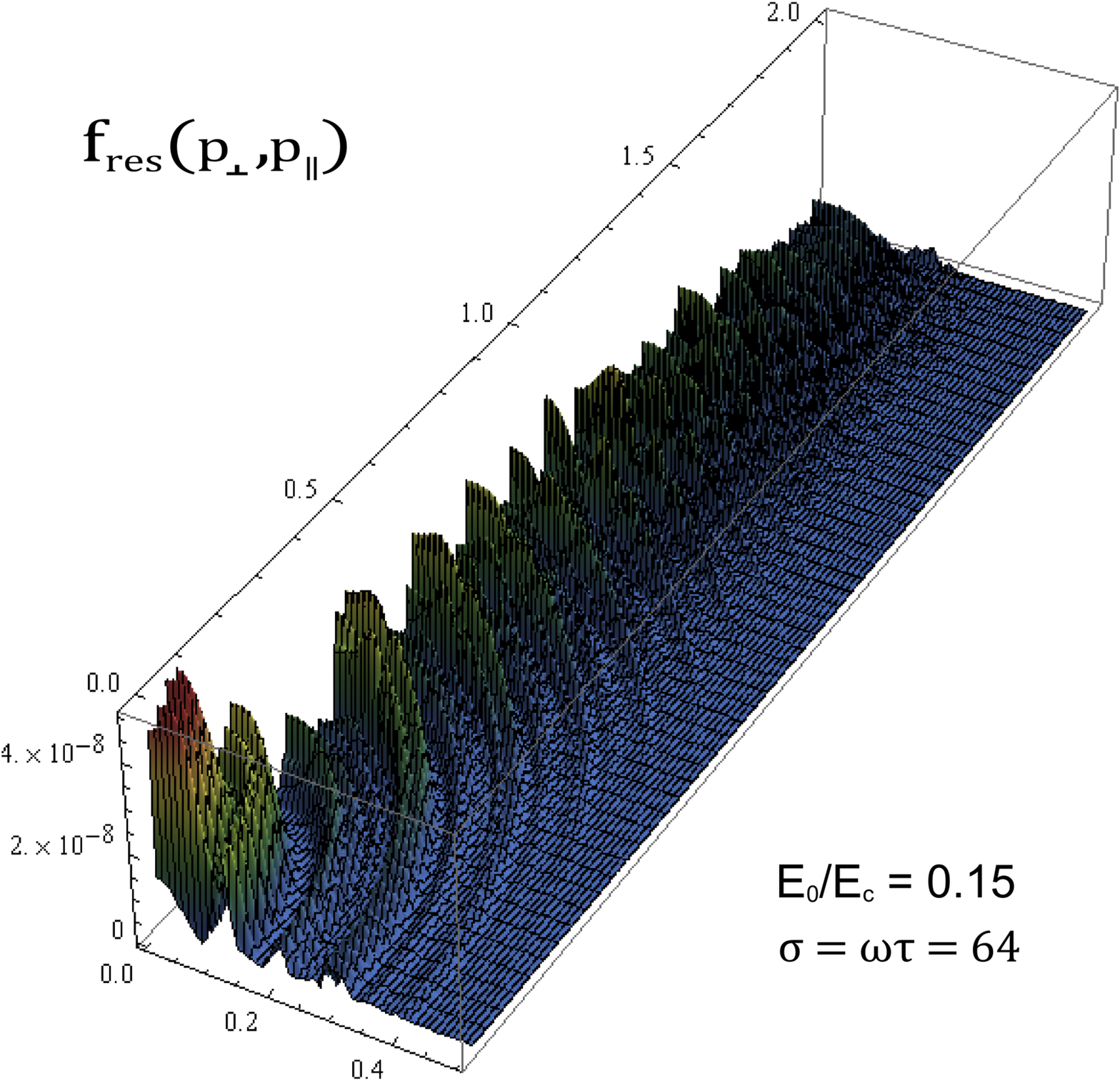}
\end{tabular}
\caption{Accumulation pattern of the distribution function for the Gaussian-modulated 
periodic field $E(t)=E_0\cos{(\omega t+\phi)}\ e^{-t^2/2\tau^2}, \quad \phi = 0$.}
\label{fig:gaussfield2}
\end{figure}

The typical picture of an EPP under the influence of the smooth Eckart field (\ref{field1}) 
is presented in Fig.~\ref{fig:eckertfield}.
The distribution function (left panel,  $p_{\bot} = p_{\parallel} = 0$) demonstrates three 
stages of the EPP evolution: quasiparticle, transient and residual.  
For the density (right panel) the fluctuations in the transient region are lost. 
The dependence of the temporal behaviour of the maximum of the distribution function on 
the field strength $E_0$ is shown in Fig.~\ref{fig:eckertfield2}.
The fluctuation region separates the quasiparticle state from the asymptotic out-state.
Since the out-state is stationary  when the external field is absent, we interprete this 
behaviour as a dynamical phase transition.
The oscillatory behaviour of the transient process is more complicated 
(see Fig.~\ref{fig:gaussfield}) 
in the case of a Gaussian-modulated  periodic field (\ref{field2}). 
In contrast to the Eckart pulse scenario, one can see oscillations of the number density. 
Our KE is of non-Markovian type, so one should expect to observe accumulation/ memory effects.
The typical picture of the accumulation case is presented in Fig.~\ref{fig:gaussfield2}. 
For long pulses (here $\sigma = 64$) the REPP distribution function has a higher value than 
for short ones (here $\sigma = 4$). 
Moreover, the oscillation maxima form a complicated pattern which is a manifestation of 
multiphoton absorption.
Similar resonance ring structures of $f_{res}(p_{\bot}, p_{\parallel})$ were obtained by 
Otto et al. \cite{Otto}.

\section{Conclusions}
The QED vacuum can be seen as nonlinear optical medium \cite{Piazza}. 
This gives us the chance to study its properties by optical methods. 
Two experimental approaches are perspective here for EPP observation: 
the registration of annihilation photons emitted from the region of EPP generation 
\cite{002,009} and the birefrigence \cite{008}. 
The existing estimations are very coarse and contradictory. 
The reason for this is that the nonlinearity of the QED vacuum makes such studies of the 
distribution function strongly dependent on magnitude, shape and duration of field pulse \cite{004}. As we have presented here, the shape of $f_{res}(p_{\bot}, p_{\parallel})$ can become 
complicated due to the non-Markovian and nonlinear nature of the system.

These facts, as detailed in the present work, underline that an important direction of  
further investigations is the search for acceptable approximate methods of the EPP 
description.

\vskip 5mm

\centerline{\bf Acknowledgment}
This work was supported by University of Wroclaw internal grant number 2467/M/IFT/14. D.B. is grateful for support  by the  Polish Ministry of Science and Higher Education (MNiSW) under grant No. 1009/S/IFT/14.
The authors would like to thank Prof. B.\ K\"ampfer (HZDR) 
for valuable discussions and for the invitation to the Kick-off 
Meeting for the Helmholtz International Beamline for Extreme Fields (HIBEF) 
at the European XFEL.

\end{document}